\documentclass[12pt]{article}
\usepackage{ssi}
\usepackage{times}

\usepackage[dvips]{graphicx}
\DeclareGraphicsExtensions{.eps}
\usepackage{psfrag}

\def\GeV{\;\mbox{GeV}}
\def\pb{\;\mbox{pb}}

\title{{\boldmath $ep$} Physics at High {\boldmath $Q^2$}}
\author{Thomas Hadig\\ 
Physikalisches Institut, Universit{\"a}t Heidelberg, 69120 Heidelberg, Germany\\
on behalf of the H1 and ZEUS Collaborations}

\begin{document}

\maketitle

\begin{abstract}%
\baselineskip 16pt

This article summarizes a talk presented at the SLAC Summer Institute
2000, SLAC, Stanford, USA.

The HERA $ep$ collider allows the measurement of the proton structure
and tests of the Standard Model in a large region of phase space that
has not been accessible before. Such tests provide the framework for the
H1 and ZEUS Collaborations to look for physics beyond the Standard
Model.

\end{abstract}

\section{Introduction}

The HERA collider provides the H1 and ZEUS detectors with electrons or
positrons at an energy of $27.5\GeV$ and protons with $920\GeV$ making
it a unique place to study the proton structure and to search for
physics beyond the Standard Model. Each experiment has collected up to
now more than $100\pb^{-1}.$ The largest fraction stems from positron
proton collisions.

In this article, processes with large momentum transfer between the
incoming lepton and the proton are described. In the next section, the
inclusive cross section and the comparison to the Standard Model is
presented. Possible extensions of the Standard Model are compared to the
data subsequently.


\section{Inclusive Cross Section}

The beam particles can interact via neutral or charged current events.
In the former, a photon or a $Z$ boson is exchanged,
\begin{equation}
  NC: e^{\pm} p \rightarrow e^{\pm} X.
\end{equation}
In deep-inelastic scattering events, the four-momentum transfered
squared $Q^2,$ is large, i.e.\ the exchanged boson is highly virtual.
The proton structure is resolved and only a fraction $x$ of the proton
momentum takes part in the scattering. The double differential cross
section can be described by
\begin{equation}
  {d^2 \sigma^{\pm}_{NC} \over dx dQ^2} =
  {2\pi\alpha^2\over x Q^4}
  \left[
   Y_+ \tilde{F}_2(x,Q^2) \mp Y_- x\tilde{F}_3(x,Q^2) - y^2 \tilde{F}_L(x,Q^2) 
  \right]
\end{equation}
with
\begin{equation}
  Y_{\pm} = 1 \pm (1-y)^2,
\end{equation}
where the plus and minus signs apply for an incoming electron
respectively positron beam.

There is a strong global dependence on $Q^{-4}$ such that processes at
high momentum transfer are strongly suppressed. The main contribution
for virtualities well below the $Z$ pole comes from the proton structure
function $F_2.$ An additional function $xF_3$ arises from parity
breaking weak interactions which enter through the $Z$ exchange and the
$\gamma Z$ interference. The last term describes the longitudinal cross
section which is a negligible contribution at high $Q^2$ and small $y.$

In charged current processes a $W^{\pm}$ boson with the charge of the
incoming lepton is exchanged,
\begin{equation}
  CC: e^{\pm} p \rightarrow \stackrel{(-)}{\nu} X.
\end{equation}

\begin{figure}[tbp]
 \begin{center}
  \includegraphics[width=\hsize,clip=]{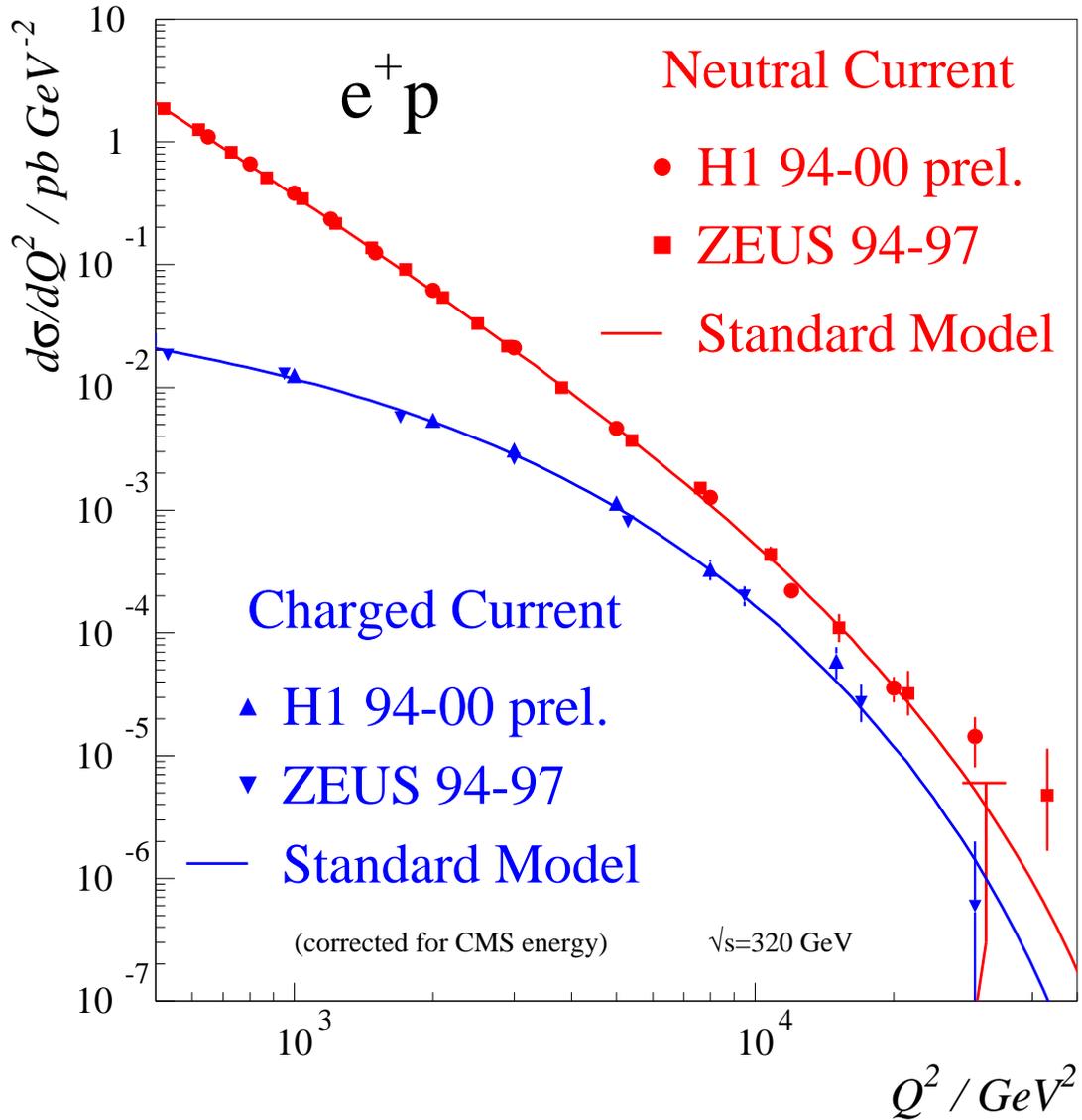}
 \end{center}
 \caption{\label{fig:ep}
  Comparison of neutral and charged current cross sections for $e^+p$ 
  collisions.
  The symbols show the data of both experiments and the full line gives
  the prediction of the Standard Model\cite{h1double,zeusnccc}.}
\end{figure}

The double differential cross section is --- in leading order --- given
by
\begin{equation}
{d^2 \sigma^{\pm}_{CC} \over dx dQ^2} =
  {G_F^2\over 2\pi x} \left( {M_W^2 \over M_W^2 + Q^2}\right)^2
  x \left[
    (u+c) + (1-y^2) (\bar{d} + \bar{s})
  \right]
\end{equation}
with the $W$ mass $M_W,$ the Fermi constant $G_F,$ and the quark
densities. At low $Q^2$ the charged current process is strongly
suppressed compared to the neutral current processes by the mass term.
This is also seen in figure~\ref{fig:ep} where the single differential
neutral and charged current cross sections are plotted.

It can be seen that the data of both experiments agree well with each
other and with the prediction of the Standard Model over more than six
orders of magnitude.

\begin{figure}[tbp]
 \begin{center}
  \includegraphics[width=\hsize,clip=]{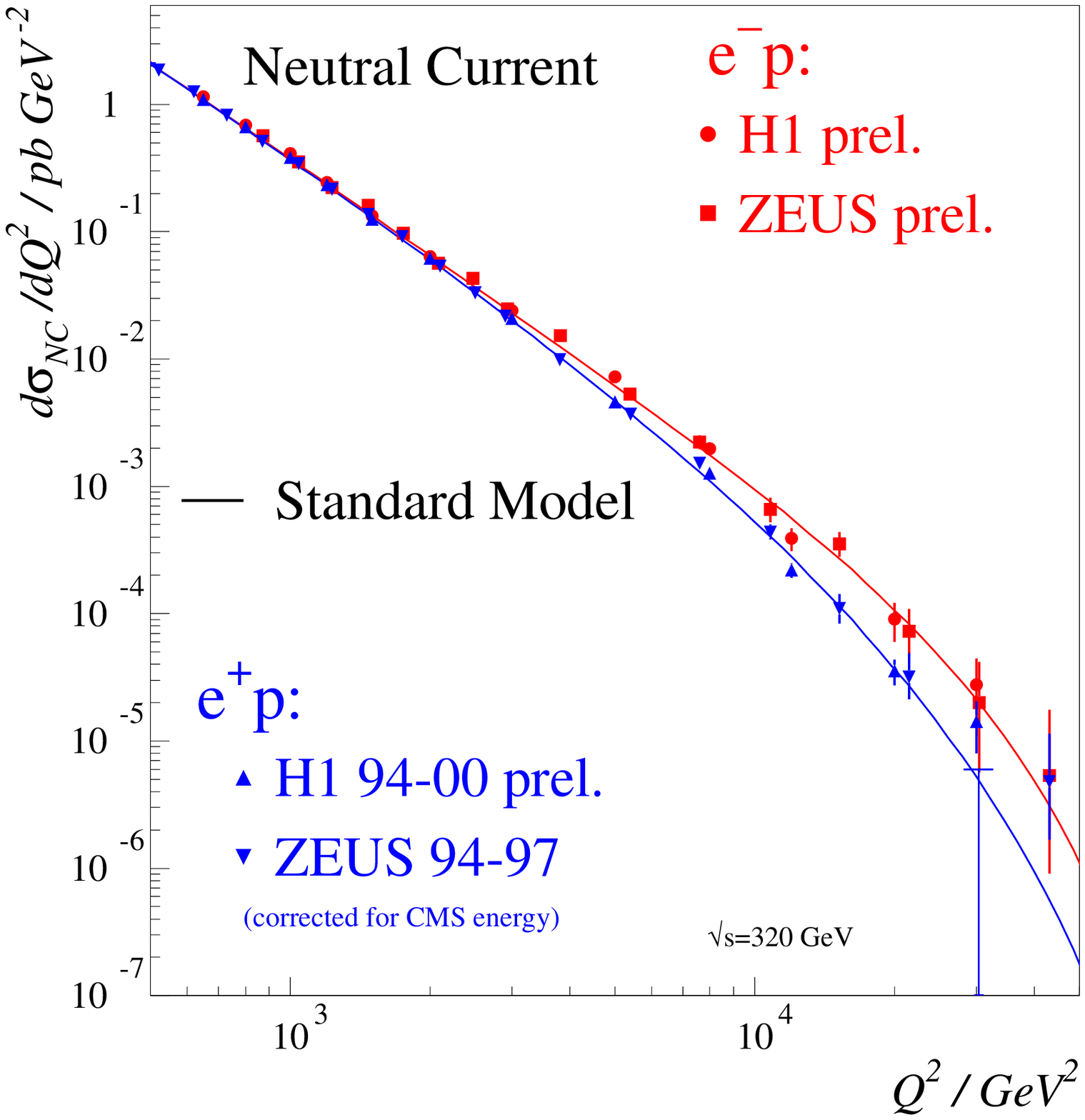}
 \end{center}
 \caption{\label{fig:nc}
  Comparison of neutral current cross sections for $e^+p$ and $e^-p$
  collisions. 
  The $e^+p$ data in the years 1994 till 1997 have been taken at a lower
  center of mass energy, a correction for the effect of this difference
  has been applied to allow for a direct comparison (see
  figure~\ref{fig:zinflu}). The symbols show the data of both
  experiments and the full line gives the prediction of the Standard
  Model\cite{zeusnccc,h1nccc,f3}.}
\end{figure}

Figure~\ref{fig:nc} shows the comparison of the neutral current cross
section for both charges of the incoming lepton. While the positron
proton data taken in the years 1994 till 1997 have been measured at a
center of mass energy of $\sqrt{s} = 300\GeV,$ the cms energy for the
other data was $\sqrt{s} = 320\GeV.$ The data from the lower energy
running has been corrected to allow a direct comparison to that from the
higher energy running.

\begin{figure}[tbp]
 \begin{center}
  \includegraphics[width=.9\hsize,clip=]{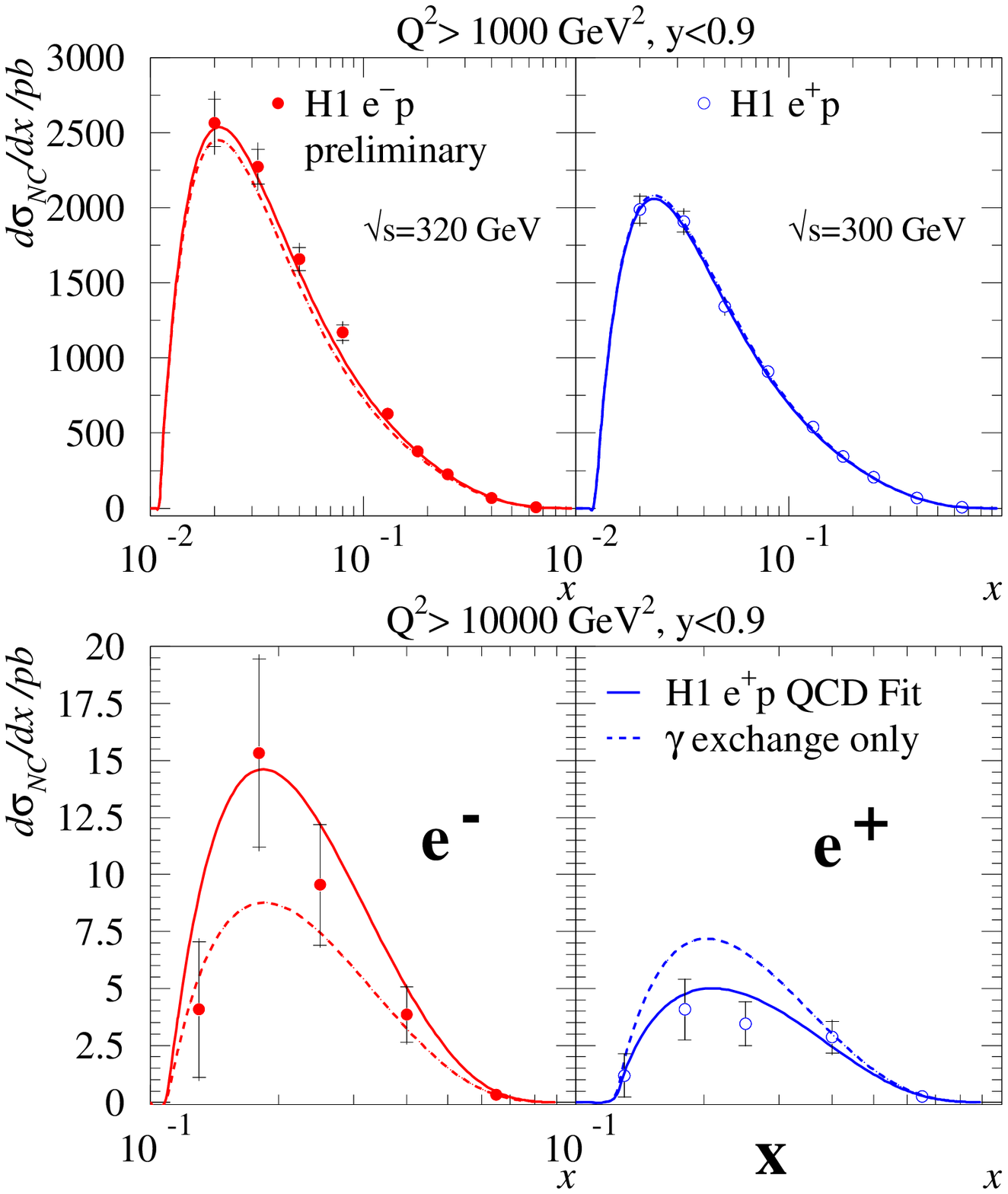}
 \end{center}
 \caption{\label{fig:zinflu}
  Influence of the $\gamma Z$ interference in two different regions of
  four-momentum transfer $Q^2.$ The symbols show the data of the H1
  Collaboration and the full line gives a fit of the Standard Model to
  the data. The dashed line shows the part of the theoretical prediction
  taking into account the exchange of photons only\cite{h1nccc}.}
\end{figure}

At low photon virtualities, the cross sections are similar, while at
higher $Q^2$ the electron cross section is significantly higher than the
one for positions. The difference is induced by the $Z$ exchange which
does not have a visible influence at low $Q^2.$ This is seen more
clearly in figure~\ref{fig:zinflu} where the single inclusive cross
section is displayed in two different regions of the momentum transfer.
In the high $Q^2$ region (bottom plots) a significant difference between
the theory predictions with and without $Z$ exchange is found. The sign
of the $\gamma Z$ interference differs for electron and positron induced
processes and the data are described well in both cases.

\begin{figure}[tbp]
 \begin{center}
  \includegraphics[width=.9\hsize,clip=]{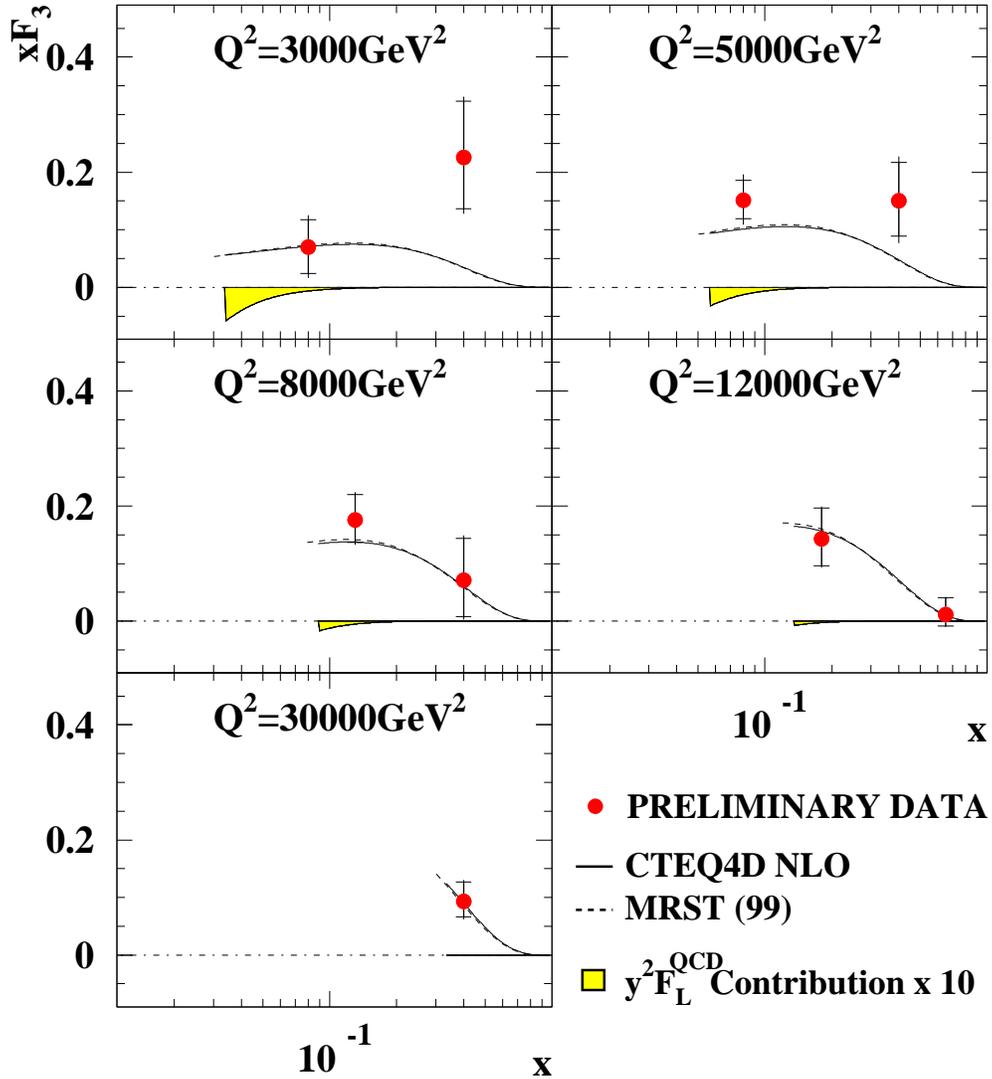}
 \end{center}
 \caption{\label{fig:f3}
  Extraction of the $xF_3$ structure function.
  The symbols show the data of the ZEUS Collaboration.
  The lines display the value taken from the global parton
  density fits performed by the CTEQ and MRS groups. The
  shaded area shows the contribution ($\times$ 10) of 
  the longitudinal structure function\cite{f3}.}
\end{figure}

This difference allows the extraction of the structure function $xF_3$
from HERA data alone as is shown in figure~\ref{fig:f3}. The data are
--- within the still large statistical errors --- consistent with those
of the global fits of the CTEQ and MRS groups. The contribution of the
longitudinal cross section in this region of phase space at large $Q^2$
and high $x$ is found to be negligible.

Detailed comparisons of data to QCD theory have been
performed. A summary of the results can be found in the talk by
J.~Engelen in these proceedings\cite{jos}.


\section{Physics Beyond the Standard Model}

One typical signature for physics beyond the Standard Model are mass
resonances in the cross section. The center of mass energy squared
available in the lepton quark cross section is
\begin{equation}
 \hat{s} = x s
\end{equation}
with the $ep$ center of mass energy $\sqrt{s} = 320\GeV.$

\begin{figure}[tbp]
 \begin{center}
  \includegraphics[width=\hsize,clip=]{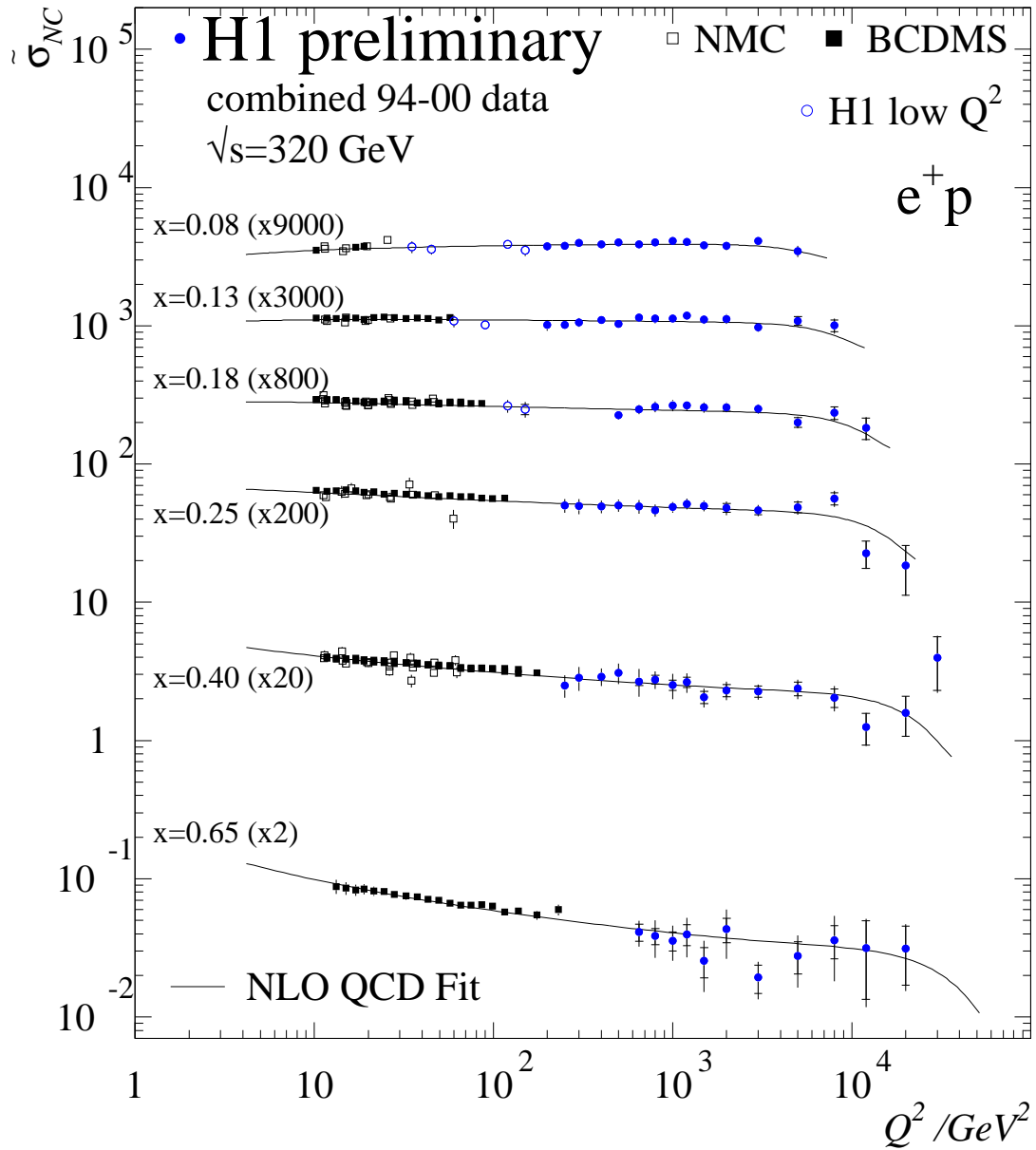}
 \end{center}
 \caption{\label{fig:doubleincl}
  Double differential inclusive cross section.
  The symbols show the data of the H1 Collaboration and the
  fixed target experiments NMC and BCDMS.
  The lines display a QCD fit to the data\cite{h1double}.
  }
\end{figure}

\begin{figure}[tbp]
 \begin{center}
  \includegraphics[width=.49\hsize,clip=]{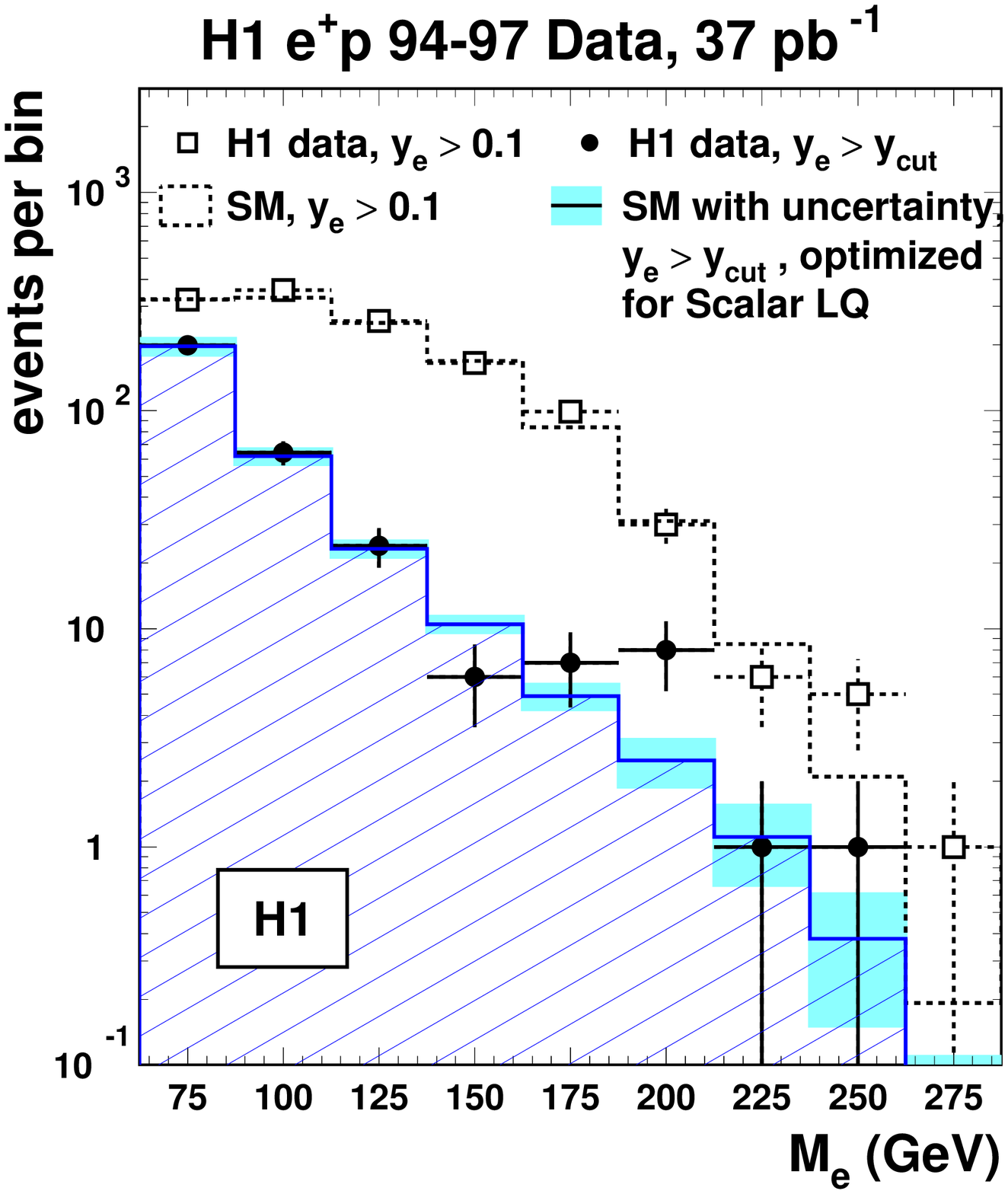}\hfil%
  \includegraphics[width=.49\hsize,clip=]{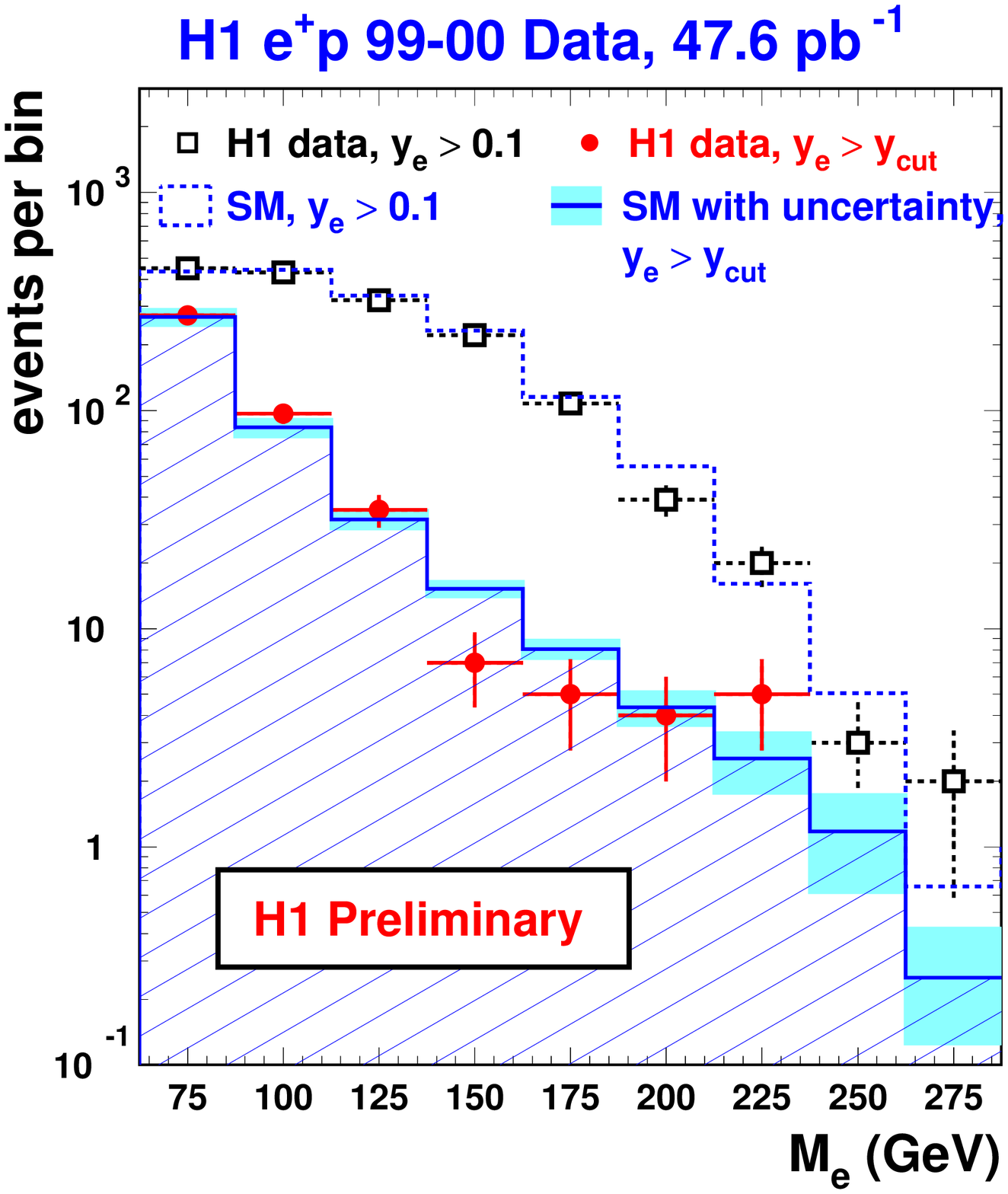}
 \end{center}
 \caption{\label{fig:leptoquark}
  Cross section as a function of the invariant mass of the lepton quark
  subprocess. Points show the data of the H1 Collaboration and the
  histograms give the Standard Model expectation. The left plots show
  the data of the 1994 to 1997 data taking, the right those of the
  1999 and 2000 data set\cite{h1lepto}. }
\end{figure}

Figure~\ref{fig:doubleincl} shows the inclusive cross section in
different bins of $x.$ The data are well described by a next-to-leading
order QCD fit. In figure~\ref{fig:leptoquark}, a more detailed study by
the H1 Collaboration is presented with cuts optimized to look for
resonances in the lepton quark cross section. While in the early data
set a deviation from the Standard Model expectation was seen, the new
data set with increased statistics does not confirm this effect. The
ZEUS Collaboration also observes, at high $Q^2$ and $x,$ no significant
deviation from the standard model expectations in their analysis of
1994-2000 data\cite{zeusosaka452}.

\begin{figure}[tbp]
 \begin{center}
  \includegraphics[width=\hsize,clip=]{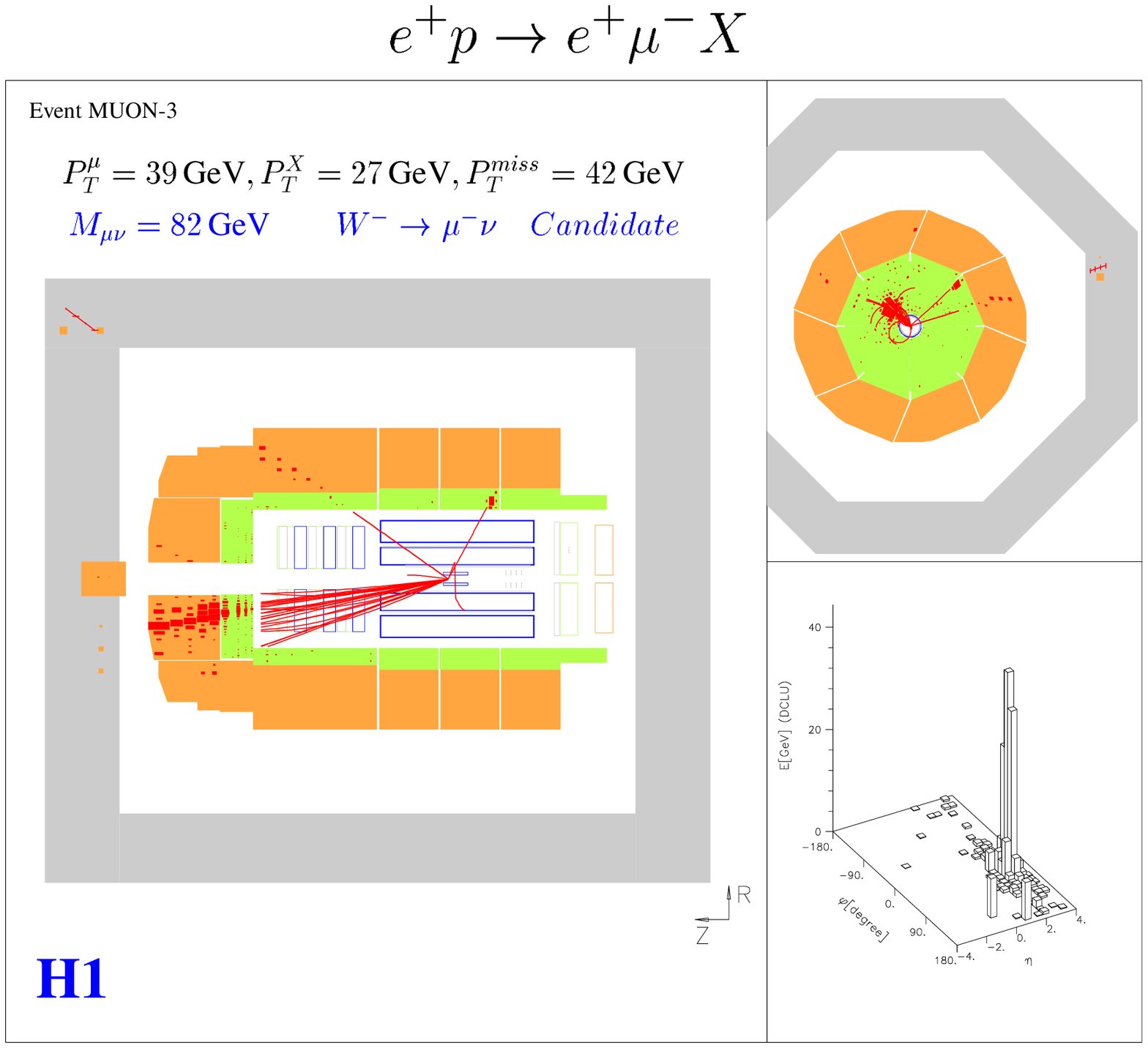}
 \end{center}
 \caption{\label{fig:isolated}
  Event display of an isolated lepton event recorded at the H1 
  detector.
  }
\end{figure}

\begin{figure}[tbp]
 \begin{center}
  \includegraphics[width=\hsize,clip=]{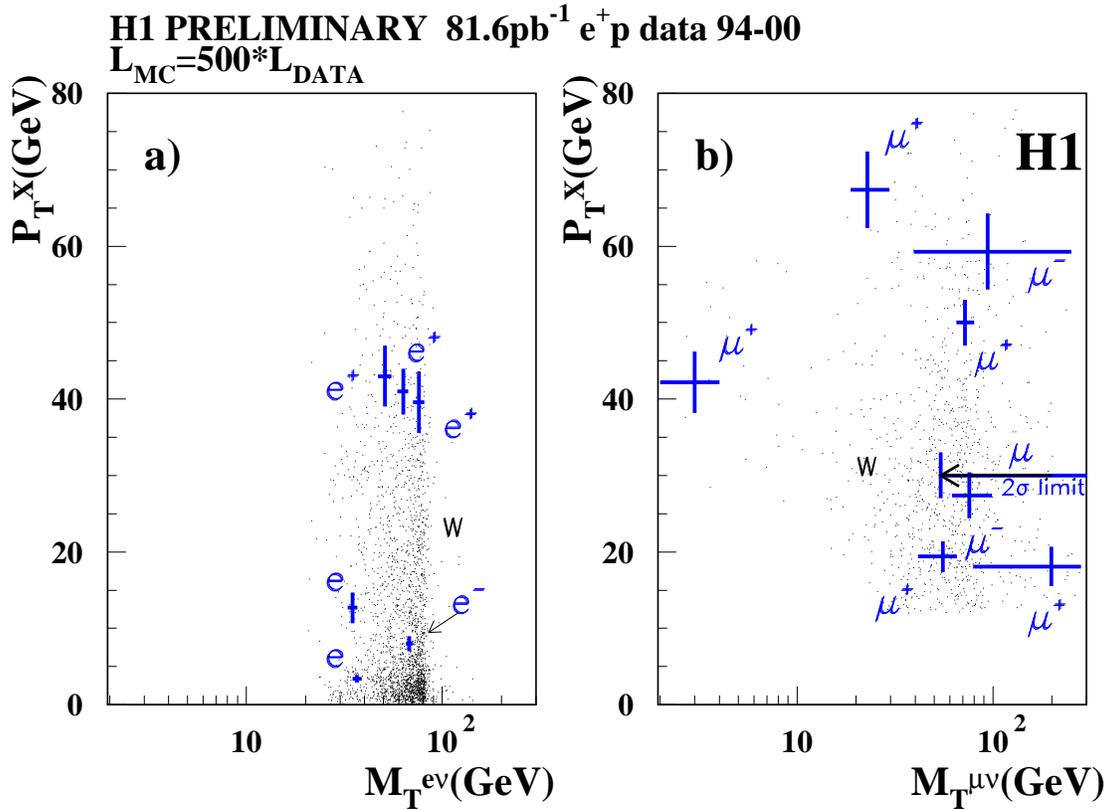}
 \end{center}
 \caption{\label{fig:masspt}
  Event distribution of the isolated lepton events. Shown is the
  transverse mass of the lepton neutrino pair versus the transverse
  momentum of the jet for isolated electron and muon events. The events
  found by the H1 Collaboration are shown as crosses corresponding to
  the uncertainty in the measurement of the observables. The small
  points show the event distribution expected from a Monte Carlo
  simulation of Standard Model $W$ production. The luminosity of the
  Monte Carlo production overshoots those of the data by a factor
  500\cite{h1iso}. }
\end{figure}

\begin{figure}[tbp]
 \begin{center}
  \includegraphics[width=.6\hsize,clip=]{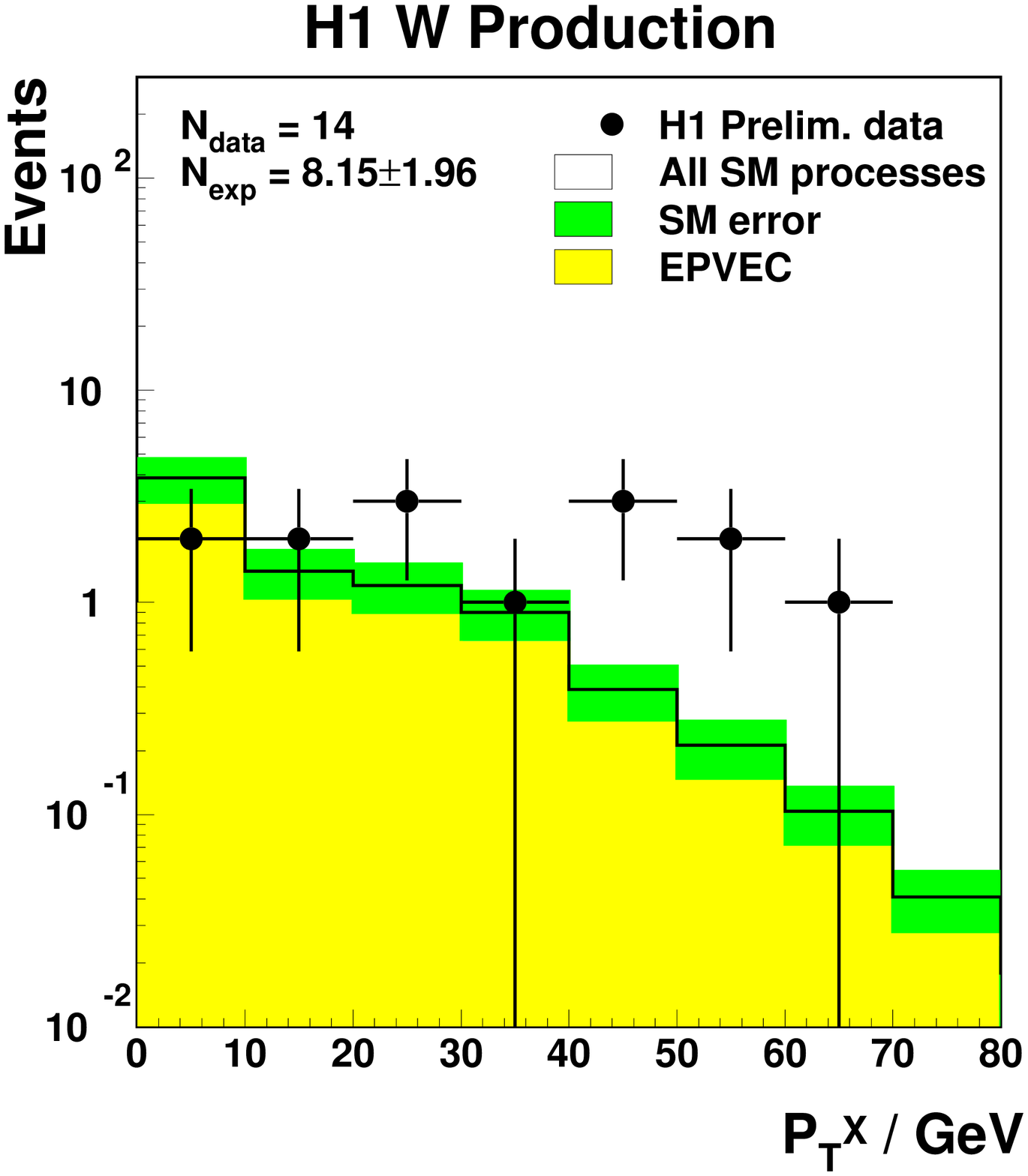}\\
  \includegraphics[width=.49\hsize,clip=]{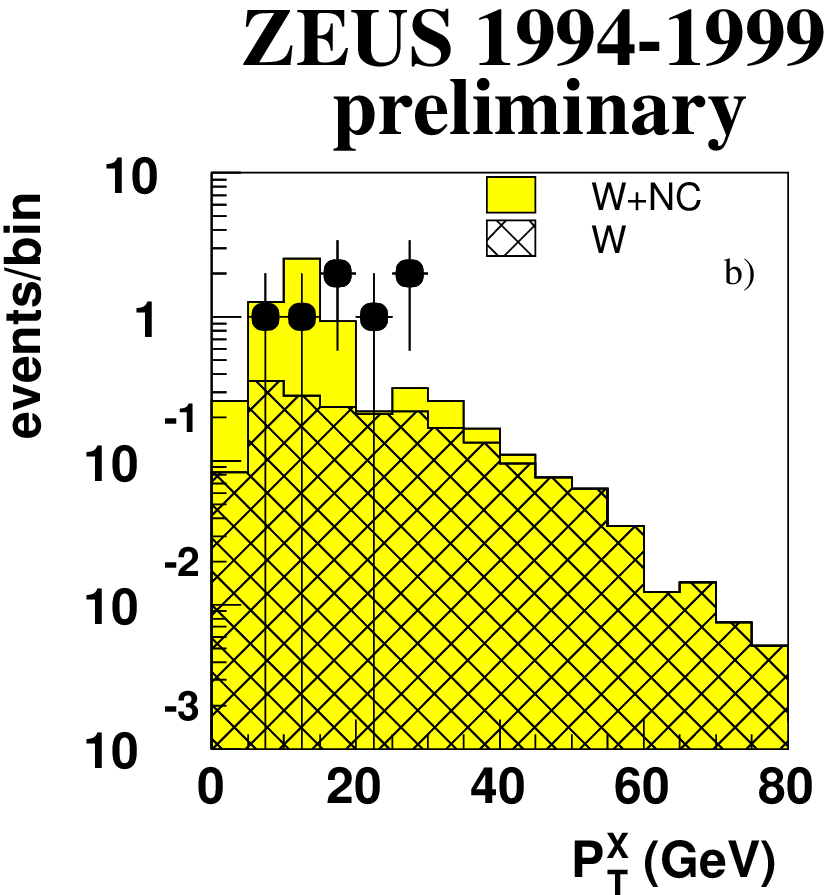}\hfil%
  \includegraphics[width=.49\hsize,clip=]{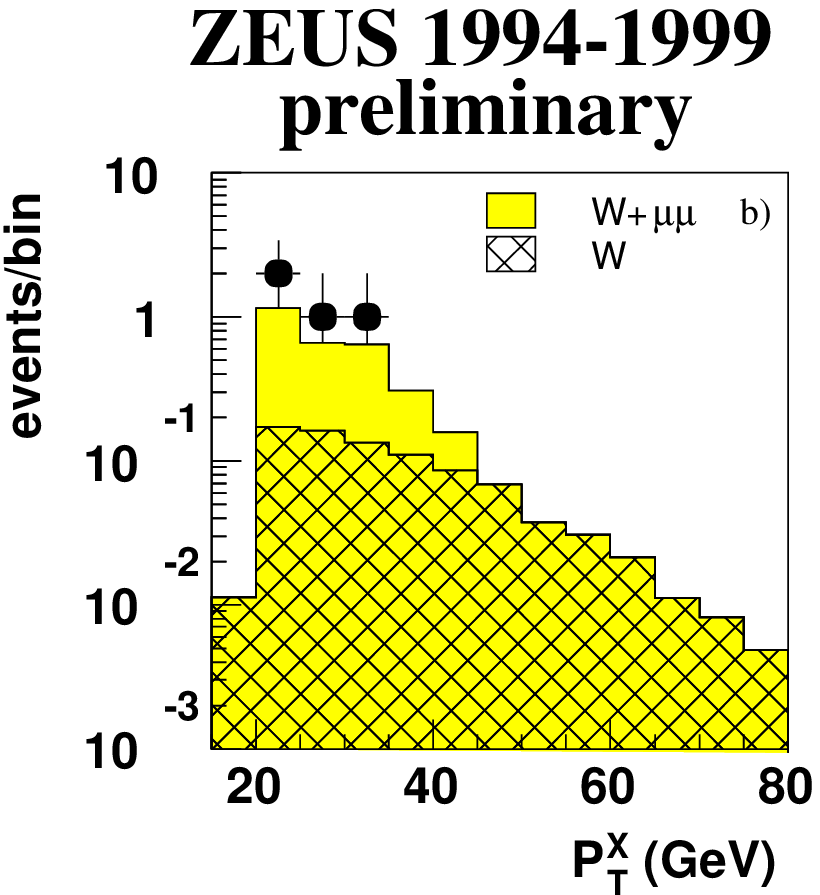}
 \end{center}
 \caption{\label{fig:pt}
  Distribution of the transverse momentum of the jets in isolated lepton
  events measured at the H1 (top) and ZEUS (bottom, left: electron
  channel, right: muon channel) Collaboration. Full points show the
  data, full lines the prediction of the Standard 
  Model\cite{h1iso,zeusiso}.}
\end{figure}

The H1 Collaboration has observed an excess in events containing
isolated leptons. A typical event is shown in figure~\ref{fig:isolated}.
The signature of the events consists of an isolated electron or muon, a
significant amount of missing transverse energy and a jet with large
transverse momentum ($p_T^X$). If the missing momentum is attributed to
a single neutrino leaving the detector undetected, the transverse mass
of the lepton neutrino pair can be extracted and is found to cluster at
the mass of the $W$ boson, see figure~\ref{fig:masspt}. However,
contrary to the distributions of the measured events, the transverse
momentum of the jet produced in Standard Model $W$ production is
predicted to be small. This is plotted in figures~\ref{fig:masspt} and
\ref{fig:pt}. While the H1 Collaboration
observes an excess at high transverse momenta of the jets, the ZEUS
Collaboration finds good agreement with the Standard Model expectations.
This picture is confirmed by the numbers given in
table~\ref{tab:lepton}. An analysis with comparable cuts shows that,
even though the Monte Carlo expectation is very similar, the H1
Collaboration sees more events than ZEUS.

\begin{table}[tbp]
 \begin{center}
  \begin{tabular}{|l|l|r|r|}
   \hline
   Collaboration & Data Set & Events seen & MC expectation \\
   \hline
   \multicolumn{4}{|c|}{Default analyses of the collaborations}\\
   \hline
   H1 preliminary & 1994-2000 $e^+p$ only, $82\pb^{-1}$ & $ 9$ & $2.3 \pm 0.6$ \\
   ZEUS prelim.   & 1994-1999 $e^+p$ and $e^-p, 82\pb^{-1}$       & $11$ & $9.8 \pm 1.3$ \\
   \hline
   \multicolumn{4}{|c|}{Special analyses using similar cuts}\\
   \hline
   H1 preliminary & 1994-2000 $e^+p$ only, $82\pb^{-1}$ & $ 9$ & $1.78$ \\
   ZEUS prelim.   & 1994-1999 $e^+p$ and $e^-p, 82\pb^{-1}$       & $ 1$ & $1.60$ \\
   \hline
  \end{tabular}
 \end{center}
 \caption{\label{tab:lepton}
  Comparison of the number of isolated lepton events seen and expected
  from Monte Carlo simulations of all relevant Standard Model processes.
  In the upper part, the default analyses of both collaborations are
  shown. Those are optimized for the detector configurations and
  specific channels. For comparison, analyses with similar cuts have
  been performed and are shown in the lower 
  part\cite{h1iso,zeusiso,talkmehta}.}
\end{table}

\begin{figure}[tbp]
 \begin{center}
  \includegraphics[width=\hsize,clip=]{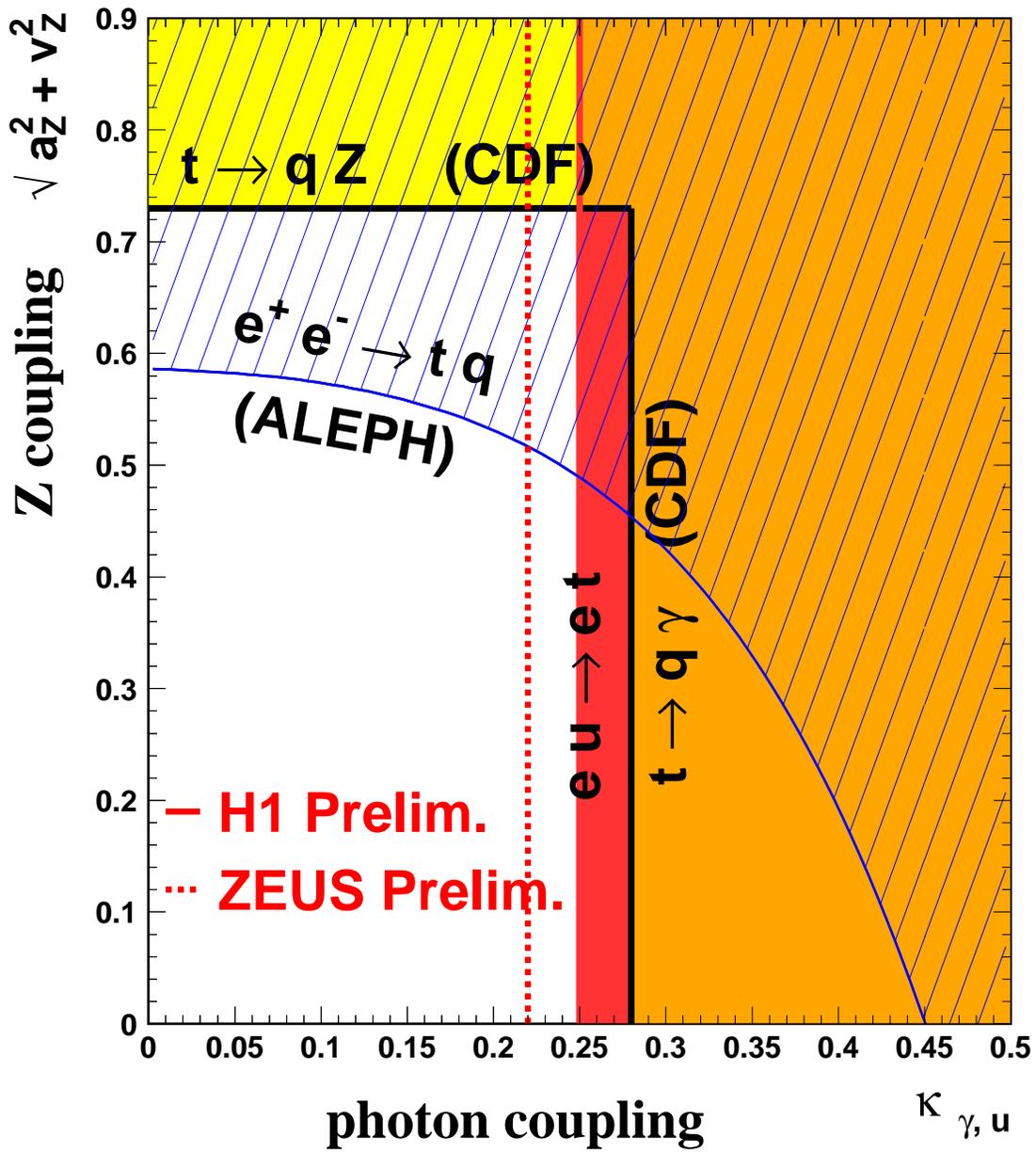}
 \end{center}
 \caption{\label{fig:top}
  Limits on the photon coupling for single top production as extracted from
  the H1 and ZEUS Collaborations. For comparison, the limits of the CDF and 
  ALEPH Collaboration are shown\cite{zeusiso,singletop}.}
\end{figure}

A possible production of single top quarks by a flavor violating neutral
current vertex has been studied by both collaborations in the leptonic
and hadronic decay channels. Since no significant excess has been found,
limits have been derived for the photon coupling. Since $Z$ exchange is
strongly suppressed, no limits on the $Z$ coupling are calculated. The
comparison in figure~\ref{fig:top} shows that the limits on the photon
coupling are the best limits available.

Super symmetric extensions of the Standard Model predict the existence
of new particles: squarks, the super symmetric partners of the quarks,
and sleptons. Standard Model particles have an R parity of +1 while SUSY
sparticles have -1. In R parity conserving theories, sparticles can only
be produced together with their anti-sparticles. In R parity violating
theories, sparticles can be produced e.g.~at lepton quark vertices where
the HERA $ep$ collider has a unique detection potential.

\begin{figure}[tbp]
 \begin{center}
  \includegraphics[width=\hsize,clip=]{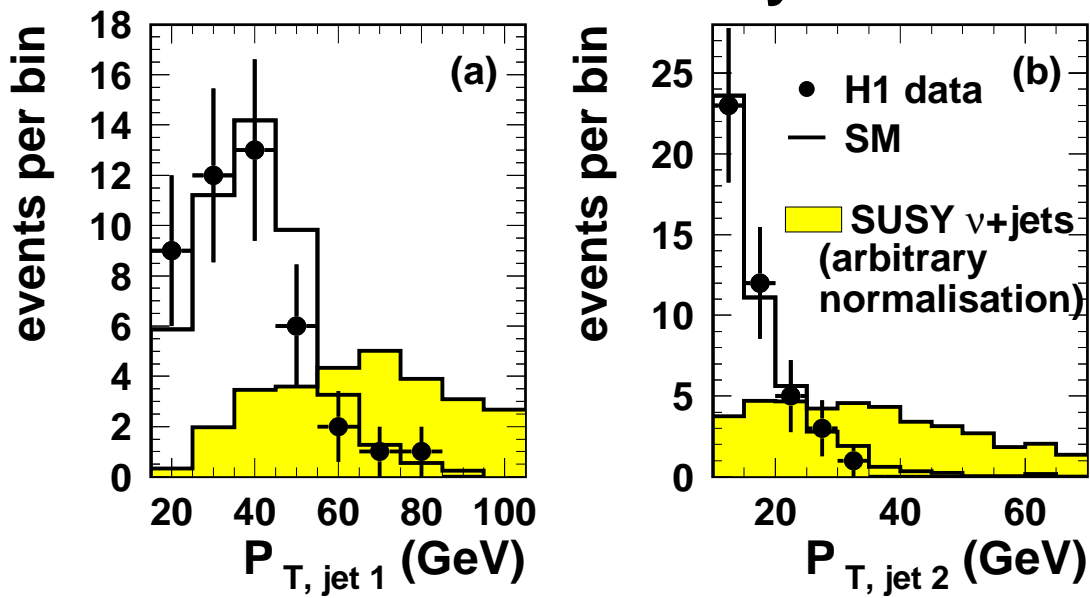}
 \end{center}
 \caption{\label{fig:susypt}
  Distribution of the transverse momentum of the jets in the neutrino
  plus multiple jet event sample. The full points show the data and the
  open histograms the expectation of the Standard Model. The filled
  histogram displays the contribution as calculated by a signal Monte
  Carlo\cite{h1rpviolsusy}.}
\end{figure}

Many decay channels for sparticles have been looked at. As an example,
the transverse momentum distribution of the jets in the neutrino plus
multiple jet channel is shown in figure~\ref{fig:susypt}.

\begin{figure}[tbp]
 \begin{center}
  \includegraphics[width=\hsize,clip=]{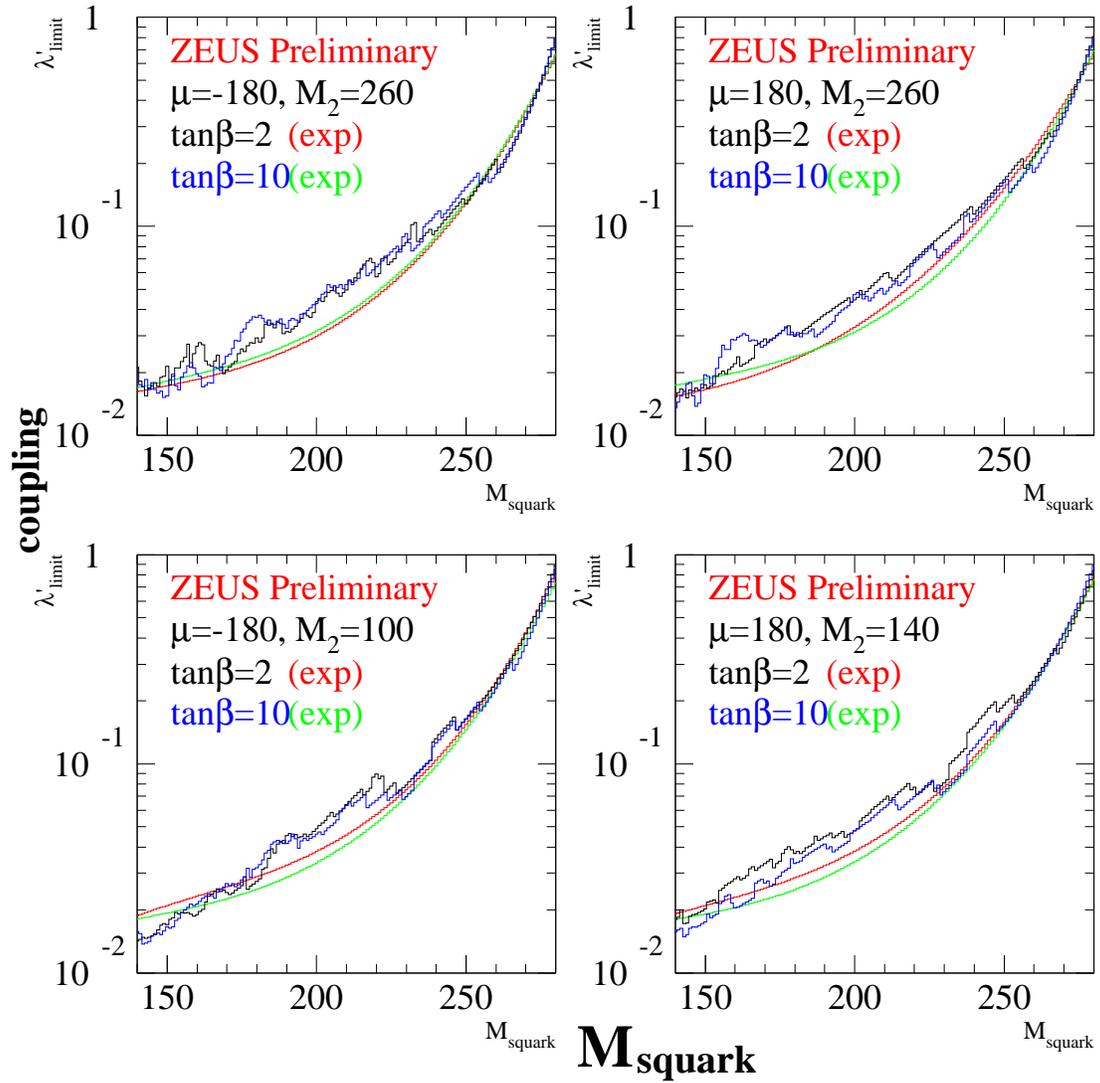}
 \end{center}
 \caption{\label{fig:susyzeus}
  95\% confidence limits for SUSY processes as extracted by the ZEUS
  Collaboration. The histograms show the limits for different
  combinations of the SUSY parameters $\mu$ and
  $M_2$\cite{zeusrpviolsusy}.}
\end{figure}

\begin{figure}[tbp]
 \begin{center}
  \includegraphics[width=\hsize,clip=]{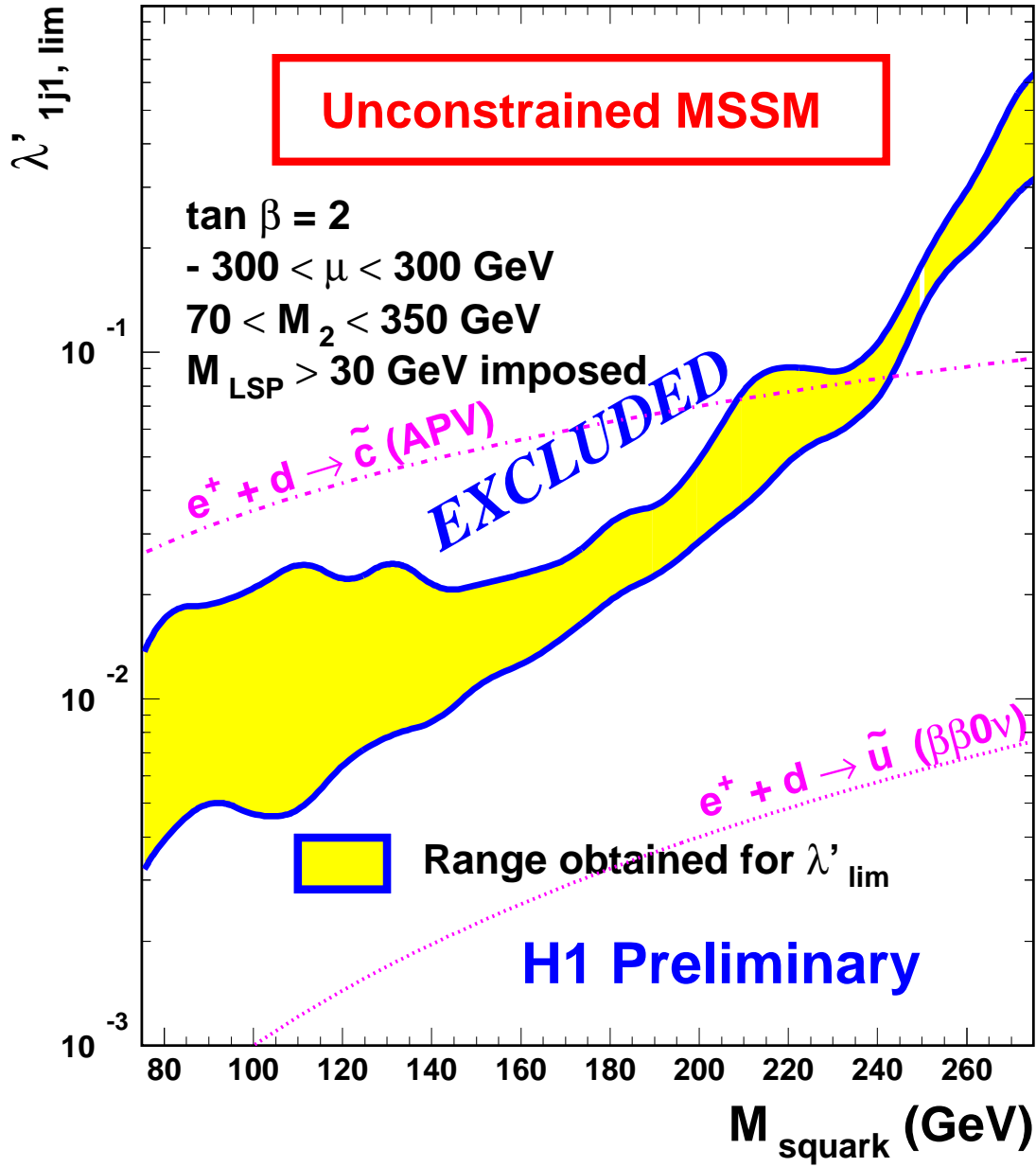}
 \end{center}
 \caption{\label{fig:susyh1}
  Limits for SUSY processes as extracted by the H1 Collaboration. The
  shaded area shows the dependence on the SUSY parameters, the region
  above the area is excluded for all parameter combinations. For
  comparison, limits from the neutrinoless double beta decay and atomic
  parity violations are shown. These are valid for first and second,
  respectively, generation squarks only, while the H1 limit is valid for
  all generation of squarks\cite{h1rpviolsusy}.}
\end{figure}

The distributions for all channels are well described by the Standard
Model and no excess is found. Therefore, limits on R parity violating
couplings have been extracted, see figures~\ref{fig:susyzeus} and
\ref{fig:susyh1}. The limits are found to depend only weakly on the 
SUSY parameters $\mu$ and $M_2$ and are the best limits available.

Additional studies have been performed and limits on e.g.~minimal
super-gravity theories\cite{h1rpviolsusy}, contact interactions, extra
dimensions\cite{contact}, or excited fermions\cite{exferm} have been
extracted.


\section{Summary}

The collider experiments at the HERA storage ring have collected a 
large amount of $ep$ data in the last eight years.

Studies of the proton structure at the H1 and ZEUS Collaborations have
shown a remarkable agreement with the Standard Model. The neutral
current as well as the charged current inclusive cross section allow the
extraction of the proton structure and tests of QCD predictions. The
influence of weak contributions is clearly seen.

The HERA $ep$ collider, also, provides a unique testing ground for
physics beyond the Standard Model. No signal was found in searches for
Leptoquarks, for single top production, and for R parity violating SUSY.
Corresponding limits have been extracted and are in large regions of the
parameter space the best limits available. The H1 Collaboration has seen
an excess of events with an isolated lepton, large missing momentum, and
a jet with large $p_t^X.$ This excess is not confirmed by the ZEUS
Collaboration. Whether these events are the first signals of physics
beyond the Standard Model or a simple statistical fluctuation can only
be decided after more data has been collected. The next data taking
period will begin after a significant upgrade of both, detectors and the
storage ring, and startup is expected for the middle of next year.




\end{document}